\documentclass[usenatbib]{mn2e}
\usepackage{amsmath}
\usepackage{amssymb}
\usepackage{graphicx}
\usepackage{times}
\usepackage[colorlinks=true,linkcolor=blue]{hyperref}

\title[Relativistic cluster models]{The dynamics of general
  relativistic isotropic stellar cluster models -- Do relativistic extensions
  of the Plummer model exist?}

\author[S. De Rijcke et al.]{S. De Rijcke$^{1}$\thanks{E-mail:
    sven.derijcke@Ugent.be}, R. Verbeke$^1$\thanks{E-mail:
    robbert.verbeke@Ugent.be}, T. Boelens$^1$\\ $^{1}$Ghent
  University, Dept. Physics \& Astronomy, Krijgslaan 281, S9, B-9000,
  Ghent, Belgium}

\begin{document}
\date{}
\pagerange{\pageref{firstpage}--\pageref{lastpage}} \pubyear{2009}

\maketitle

\label{firstpage}

\begin{abstract}
We show that the general relativistic theory of the dynamics of
isotropic stellar clusters can be developed essentially along the same
lines as the Newtonian theory. We prove that the distribution function
can be derived from any isotropic momentum moment and that every
higher-order moment of the distribution can be written as an integral
over a zeroth-order moment. 

We propose a mathematically simple expression for the distribution
function of a family of isotropic general relativistic cluster models
and investigate their dynamical properties. In the Newtonian limit,
these models obtain a distribution function of the form $F(E) \propto
(E-E_0)^\alpha$, with $E$ binding energy and $E_0$ a constant that
determines the model's outer radius. The slope $\alpha$ sets the
steepness of the distribution function and the corresponding radial
density and pressure profiles. We show that the field equations only
yield solutions with finite mass for $\alpha \le 3.5$. Moreover, in
the limit $\alpha \rightarrow 3.5$, only Newtonian models exist. In
other words: within the context of this family of models, no general
relativistic version of the Plummer model exists. The most strongly
bound model within the family is characterized by $\alpha=2.75$ and a
central redshift $z_c \approx 0.55$.
\end{abstract}

\begin{keywords}
galaxies: kinematics and dynamics -- galaxies: nuclei -- physical data
and processes: relativistic processes
\end{keywords}
\newcommand{\beqn}{\begin{equation}}
\newcommand{\neqn}{ \end{equation}}

\section{Introduction}\label{intro}

The so-called Plummer model was introduced by \citet{p11} as a
description of the stellar density distribution in Galactic globular
clusters \citep{p11}. Subsequently, \citet{e16} showed that this
spherically symmetric density profile could be derived from a
phase-space distribution of the form
\begin{equation}
F(\epsilon) \propto (-\epsilon)^{7/2},
\end{equation}
with $\epsilon = \psi+v^2/2 $ the Newtonian specific energy of a star
and $\psi$ the Newtonian gravitational potential of the stellar
cluster \citep{de87}. This distribution function self-consistently
generates a mass distribution with a gravitational potential \beqn
\psi(r) = -\frac{GM}{a} \frac{1}{\sqrt{1+\left( \frac{r}{a}
    \right)^2}} \neqn and density profile \beqn \rho(r) =
\frac{3}{4\pi} \left( 1+\left( \frac{r}{a}
\right)^2\right)^{-\frac{5}{2}}\frac{M}{a^3} = \frac{3}{4\pi} \left(
-\frac{a}{GM} \psi \right)^5\frac{M}{a^3}.  \neqn Here, $M$ is the
total mass of the cluster and $a$ a scale length.

Certain general relativistic (GR) extensions of the Plummer model can
already be found in the literature  and we give an overview here.
For instance, in the case of a spherically symmetric cluster, the
density, the potential (or some generalization thereof), and the
distribution function are all functions of one argument, so it makes
sense to construct the metric around a single, unknown function of the
radius, usually denoted simply by $f(r)$. An example, inspired by the
Schwarzschild metric is \beqn ds^2 = \left( \frac{1-f}{1+f} \right)^2
c^2 dt^2 - (1+f)^4( dr^2 + r^2 d\Omega^2 ).  \neqn In the Newtonian
limit, $f(r)$ reduces to $-\psi/(2c^2)$.

One approach is to choose $f(r)$ such that it produces a meaningful
cluster model in the Newtonian limit, for instance by equating it to
the gravitational potential of the Newtonian cluster. \citet{nl13}
show how this technique can be used to recover GR extensions of the
hypervirial models of which the Plummer model is a special
case. Solving the time-time-component of the field equations yields a
density that together with $f$, by construction, correctly reduces to
the corresponding Newtonian potential-density pair. However, as these
authors note, the pressure does not reduce to the expected Newtonian
limit. This is because the underlying distribution function does not
reduce to the proper Newtonian limit.

Another possibility is to equate the radial and tangential field
equations, thus enforcing isotropy, and to solve the resulting
equation for the metric. This solution can then be plugged in the
time-time-component of the field equations to yield the density
profile. \citet{bu64} has used this procedure to produce a cluster
model with an equation of state analogous to that of the Plummer
model, i.e. a polytrope with index $n=5$. \citet{fa71}, using a metric
of the form \beqn ds^2 = e^{\nu(r)} c^2 dt^2 -e^{\lambda(r)} dr^2 -
r^2 d\Omega^2, \neqn subsequently derived a rather unwieldy analytical
expression for the distribution function of this model and showed
that, unless the central value of the potential satisfies
$\exp(\nu(0)) > 0.413$, it can show a ``temperature inversion'' in
the sense that it is not a monotonically decreasing function of
energy. Alternatively, one can impose the polytropic equation of state
on the field equations, which yields a generalization of the
Lane-Emden equation, and thus solve for the unknown function in the
metric \citep{to64,ka67}.

Both techniques avoid an explit calculation of the distribution
function. Using a generalization of Eddington's integral equation, it
can, however, be determined from the density profile
\citep{fa68,pk96}. Unfortunately, this distribution function is not
guaranteed to be positive everywhere in phase space, although
necessary conditions for positivity have been derived \citep{su77}.

Since employing the Eddington integral equation can lead to rather
cumbersome expressions for the distribution function and, moreover,
the latter's positivity is not guaranteed from the outset, we here
advocate another approach. We first write down a mathematically simple
distribution function that is everywhere positive and that reduces to
a well-defined Newtonian limit. From this distribution function, the
density and pressure profiles can be calculated. By construction, all
moments of the distribution function will reduce to the proper
Newtonian limit. Solving the field equations finally yields the
metric. While such generalizations of Newtonian cluster models may not
have the same equation of state as in the Newtonian limit, they have
the benefit of having a mathematically simple, strictly non-negative
distribution function with a properly defined, meaningful Newtonian
limit. Our goal is to produce general relativistic cluster models with
isotropic, polytropic distribution functions, to study their dynamical
properties, and to investigate their Newtonian limits. In particular,
we wish to study how the Newtonian polytropes, of which the well-known
Plummer model is a special case, fit in this more general scheme of
models.

In section \ref{isodyn}, we develop the dynamical theory of general
relativistic stellar cluster models and calculate the properties of
models with isotropic, polytropic distribution functions. In section
\ref{solve}, we present our method for solving the field equations for
such models. We end with a discussion of the models in section
\ref{discussion} and conclude in section \ref{conc}.

\section{Isotropic dynamical models for general relativistic stellar clusters}
\label{isodyn}

\subsection{The internal dynamics of isotropic clusters}

In general relativistic dynamics, the distribution function (DF)
$F(x^\mu,p^i) d^3x d^3p$ counts the number of occupied world lines
that intersect a 6-dimensional submanifold of the 8-dimensional phase
space. This 6-dimensional submanifold consists of a 3-dimensional
spatial hypersurface and its future mass hyperboloid. In the absence
of particle creation/annihilation or collisions, the Lie derivative of
the DF is zero, or \beqn \left[ p^\mu \frac{\partial}{\partial x^\mu}
  - \Gamma^\mu_{\alpha\beta}p^\alpha p^\beta \frac{\partial}{\partial
    p^\mu} \right] F(x,\vec{p}) = 0.  \neqn 

Let $\hat{p}^\mu$ be the components of the momentum 4-vector in a
local orthonormal frame at rest, such that $\hat{p}_0^2 - \sum_i
\hat{p}_i^2 = (mc)^2$, with $m$ the rest mass of a single star. In
such a local orthonormal frame, tensor quantities of the form \beqn
T_{\mu \nu \ldots \kappa}(x) = \int \frac{\hat{p}_\mu \hat{p}_\nu \ldots
  \hat{p}_\kappa}{\hat{p}_0} F(x,\vec{p}) d\hat{p}_1 d\hat{p}_2
d\hat{p}_3 \neqn can be defined. If the Lie derivative of the DF
disappears, then all these quantities have zero covariant
divergence. The most well-known such tensor quantities are those with
one index (the stream density vector) and two indices (the
energy-momentum tensor).

In an isotropic cluster, the DF depends only on $p_0$, the 0-component
of the momentum 4-vector, which is a constant in a time-independent
gravitating system (see below). Obviously, what matters in the above
definition of the momentum moments of the DF is the number of
instances of each momentum component. We therefore re-write these
momentum moments as \beqn \mu_{k,2m,2n,2l}(x) = \int 
  {\hat{p}_0}^k \hat{p}_1^{2m} \hat{p}_2^{2n}
  \hat{p}_3^{2l} F(p_0) \frac{d\hat{p}_1 d\hat{p}_2
d\hat{p}_3}{\hat{p}_0}.  \neqn Using the parameterization
\begin{align}
\hat{p}_0 &= \sqrt{ (mc)^2+p^2 } \nonumber \\
\hat{p}_1 &= p \cos \vartheta \nonumber \\
\hat{p}_2 &= p \sin \vartheta \cos \varphi \nonumber \\
\hat{p}_3 &= p \sin \vartheta \sin \varphi
\end{align}
this reduces to
\begin{align} \mu_{k,2m,2n,2l}(x) &= \frac{1}{2\pi} \frac{
  \Gamma\left(m+\frac{1}{2}\right) \Gamma\left(n+\frac{1}{2}\right)
  \Gamma\left(l+\frac{1}{2}\right)}{
    \Gamma\left(m+n+l+\frac{3}{2}\right) } \times \nonumber
  \\ & \hspace*{2em} 4\pi \int F(p_0) \hat{p}_0^{k-1} p^{2(m+n+l)+2} dp.
\end{align}
Let $E$ be the energy of a star as measured by an obsever at rest at
infinity, where the geometry of spacetime is essentially flat. The
energy measured by a local observer at rest, denoted by $E_{\sf
  local}$, is linked to $E$ via \beqn E = \sqrt{g_{00}} E_{\sf local}
= e^{\phi/2} E_{\sf local} = cp_0. \neqn The 0-momentum in the local
orthonormal frame, $\hat{p}_0$, is related to the energy at infinity
as \beqn  c \hat{p}_0 = \frac{c p_0 }{\sqrt{g_{00}}} =
\frac{E}{\sqrt{g_{00}}}.
\neqn
Therefore, 
\beqn
E^2 - m^2 c^4 g_{00} = g_{00} p^2 c^2
\neqn
and 
\beqn
p dp = \frac{E dE}{c^2 g_{00}}.
\neqn
Then
\begin{align}
\mu_{k,2m,2n,2l}(x) &= \frac{1}{2\pi} \frac{
  \Gamma\left(m+\frac{1}{2}\right) \Gamma\left(n+\frac{1}{2}\right)
  \Gamma\left(l+\frac{1}{2}\right)}{
  \Gamma\left(m+n+l+\frac{3}{2}\right) } \times 
\nonumber \\ & \hspace*{2em} \mu_{k,2(m+n+l)}(x)
\end{align}
which defines the set of isotropic $k$-moments \begin{align}
  \mu_{k,2q} &= 4 \pi \int F(p_0) \hat{p}_0^k \left(
  \hat{p}_0^2-m^2c^2 \right)^{q+\frac{1}{2}} d\hat{p}_0 \nonumber
  \\ &= \frac{(mc)^{2q+k+2}}{2^{q+\frac{3}{2}} E_0^{2q+k+2}}
  \tilde{\mu}_{k,2q}(E_0^2) \label{mu2q}
\end{align}
with \begin{align} \tilde{\mu}_{k,2q}(E_0^2) &= \nonumber
  \\ & \hspace*{-2em} 2^{q+\frac{5}{2}}\pi \int_{E_0^2} F(E^2) \left(
  E^2 \right)^{\frac{k-1}{2}} (E^2-E_0^2)^{q+\frac{1}{2}}
  dE^2 \label{mut2q} \end{align} and $E_0^2 = (mc^2)^2
g_{00}$. Deriving this equation $q$ times with respect to $E_0^2$
leads to \beqn \tilde{\mu}_{k,0}(E_0^2) = \frac{(-1)^q}{(2q+1)!!}
D^q_{E_0^2} \tilde{\mu}_{k,2q}(E_0^2).  \neqn Here, \beqn (2q+1)!! =
(2q+1)(2q-1) \ldots 1 \neqn indicates the double factorial. This
equation is formally identical to equation (1.3.7) in \citet{de86}. We
can therefore simply invoke equation (1.3.8) from that same work to
invert the above expression and to write all higher order $k$-moments
of the DF in terms of the zeroth-order $k$-moment: \beqn
\tilde{\mu}_{k,2q}(E_0^2) = \frac{(2q+1)!!}{(q-1)!!} \int_{E_0^2}
(E^2-E_0^2)^{q-1} \tilde{\mu}_{k,0}(E^2) dE^2.  \neqn

With the aid of equation (1.3.12) from \citet{de86}, (\ref{mut2q}) can
be inverted as
\begin{align} 
E^{k-1} F(E) & \nonumber \\
&\hspace*{-4em}= \frac{1}{2^{q+\frac{5}{2}} \pi^{\frac{3}{2}} \Gamma\left(
  q+\frac{3}{2}\right)} D^{q+2}_{E^2} \int_{E^2}
\frac{\tilde{\mu}_{k,2q}({E}_0^2)}{\sqrt{ {E}_0^2-E^2}} d{E}_0^2 \nonumber \\
&\hspace*{-4em}= \frac{(mc)^{-2q-k-2}}{2 \pi^{\frac{3}{2}} \Gamma\left(
  q+\frac{3}{2}\right) } D^{q+2}_{E^2} \int_{E^2}
\frac{E_0^{2q+k+2}\mu_{k,2q}({E}_0^2)}{\sqrt{ {E}_0^2-E^2}} d{E}_0^2. 
\end{align}
In particular, for $(q=0,k=2)$ and for $(q=1,k=0)$ the above inversion
relation reduces to the two important special cases
\begin{align} E F(E) &=
\frac{1}{\pi^2 m^4c^3} D^2_{E^2} \int_{E^2} \frac{E_0^4\rho(E_0^2)}{\sqrt{
    E_0^2-E^2}} dE_0^2 \nonumber \\ \frac{1}{E}F(E) &= \frac{2}{\pi^2
  m^4 c^5} D^3_{E^2} \int_{E^2} \frac{E_0^4P(E_0^2)}{\sqrt{
      E_0^2-E^2}} dE_0^2
\end{align}
with $\rho$ the mass density and $P$ the pressure. These are none
other than the inversion relations derived by \citet{fa68} and
\citet{pk96}. Here, we made use of the fact that
\begin{align}
\mu_{2,0} &= \frac{4\pi}{g_{00}^2c^4} \int F(E) E^2 \sqrt{ E^2-E_0^2 } dE = \rho c \nonumber \\
\mu_{0,2} &= \frac{4\pi}{g_{00}^2c^4} \int F(E) \left( E^2-E_0^2\right)^{\frac{3}{2}} dE = \frac{3P}{c}
\end{align}
\citep{zp65,os74}. Hence, we have shown that these two relations
linking the density and pressure to the isotropic DF are simply
specific cases of a more general link between the DF and any of its
moments $\mu_{k,2q}$.

\subsection{A generalized polytropic distribution function}

For a static, spherically symmetric gravitational system, the metric
can always be brought in the form \beqn ds^2 = e^{\phi(r)} c^2 dt^2 -
\left( 1-\frac{2GM(r)}{c^2r} \right)^{-1} dr^2 - r^2 d\Omega^2, \neqn
with $M(r)$ the total gravitating mass interior to the areal radius
$r$ and $\phi$ a potential function that, in the Newtonian limit,
reduces to $2\psi/c^2$. We propose a distribution function of the form
\beqn F(E)= f_0 \left( \frac{mc^2}{E} \right)^{2\beta}\left( \frac{
  m^2c^4 e^{\Phi}-E^2}{m^2c^2} \right)^\alpha, \label{propdf} \neqn with $\alpha$ and
$\beta$ positive real numbers, $f_0$ a constant forefactor, and $\Phi
= \phi(R)$, the value of the potential at the outer edge of the
cluster at radius $r=R$.

For the isotropic distribution function given above, the energy
density is given by
\begin{align}
\rho c^2 &= \frac{4\pi}{c^3} e^{-2\phi}
\int_{mc^2e^{\phi/2}}^{mc^2e^{\Phi/2}} F(E) E^2 \sqrt{ E^2 - m^2c^4
  e^\phi} dE \nonumber \\ &= \pi^{3/2}
\frac{\Gamma(\alpha+1)}{\Gamma\left(\alpha+\frac{5}{2}\right)} f_0 m^4
c^{3+2\alpha} e^{-2\phi} \left( e^\Phi-e^\phi \right)^{\alpha+\frac{3}{2}}
\times \nonumber \\ & e^{\left(\frac{1}{2}-\beta \right)\Phi}
       {_2F_1}\left(\beta-\frac{1}{2},\alpha+1;\alpha+\frac{5}{2}; \frac{e^\Phi-e^\phi}{e^\Phi} \right)
\end{align}
\citep{zp65,os74}. Here, $\Gamma(x)$ is Euler's gamma-function and
${_2F_1}(a,b;c;z)$ is the Gaussian hypergeometric function
\begin{align}
{_2F_1}(a,b;c;z) &= \frac{\Gamma(c)}{\Gamma(b)\Gamma(c-b)} \int_0^1 t^{b-1}
(1-t)^{c-b-1}\frac{dt}{(1-zt)^a} \nonumber \\
&= \sum_{n\ge 0} \frac{(a)_n(b)_n}{(c)_n} \frac{z^n}{n!}
\end{align}
with $(q)_n$ the Pocchammer symbol, defined as $(q)_n =
\Gamma(q+n)/\Gamma(q)$.  We can choose a scale-length $a$ and denote
the scaled radius by $x = r/a$. With the choice of a mass scale $M$,
we can introduce the dimensionless parameter \beqn {\cal A}=
\frac{c^2a}{2GM}. \label{Qparam} \neqn If the mass scale $M$ is taken
to coincide the model's total mass, then $\cal A$ is simply the ratio
of the scale-length $a$ to the model's Schwarzschild radius.
We can then take \beqn f_0 =
\frac{3}{4\pi^{\frac{5}{2}}c^{3+2\alpha}}
\frac{\Gamma\left(\alpha+\frac{5}{2}\right)}{\Gamma(\alpha+1)}
\frac{M{\cal A}^{\alpha+\frac{3}{2}}}{m^4 a^3}.  \neqn With this choice for
the forefactor $f_0$, we find the following expression for the density
\begin{align}
 \rho(\phi) &= \frac{3}{4\pi} {\cal A}^{\alpha+\frac{3}{2}}
 e^{-2\phi} \left( e^\Phi-e^\phi \right)^{\alpha+\frac{3}{2}} \times \nonumber
 \\ & e^{\left(\frac{1}{2}-\beta \right)\Phi}
    {_2F_1}\left(\beta-\frac{1}{2},\alpha+1;\alpha+\frac{5}{2};
    \frac{e^\Phi-e^\phi}{e^\Phi} \right)\frac{M}{a^3}.
\end{align}
Clearly, the choice $\beta=0.5$ yields the ``simplest'' mass density
profile since in that case the hypergeometric function is identically
one and
\beqn
 \rho(\phi) = \frac{3}{4\pi} {\cal A}^{\alpha+\frac{3}{2}}
 e^{-2\phi} \left( e^\Phi-e^\phi \right)^{\alpha+\frac{3}{2}}\frac{M}{a^3}.
\neqn

The expression for the pressure follows
from
\begin{align}
P &= \frac{4\pi}{3 c^3 } e^{-2\phi}
\int_{mc^2e^{\phi/2}}^{mc^2e^{\Phi/2}} F(E) \left( E^2 - (mc^2)^2
e^\phi \right)^{\frac{3}{2}} dE \nonumber \\ &= \frac{3}{4\pi}
\frac{{\cal A}^{\alpha+\frac{3}{2}}}{(2\alpha+5)}
e^{-\left(\beta+\frac{1}{2}\right)\Phi} e^{-2\phi} \left(
e^\Phi-e^\phi \right)^{\alpha+\frac{5}{2}} \nonumber \\ & \hspace{3em} \times
{_2F_1}\left( \beta+\frac{1}{2}, \alpha+1; \alpha+\frac{7}{2}; \frac{e^\Phi-e^\phi}{e^\Phi}
\right)\frac{Mc^2}{a^3}.
\end{align}

The proper mass density is given by
\begin{align}
nm &= \frac{4\pi m}{c^3} e^{-3\phi/2}
\int_{mc^2e^{\phi/2}}^{mc^2e^{\Phi/2}} F(E) E \sqrt{ E^2 - m^2c^4
  e^\phi} dE \nonumber \\ 
&= \frac{3}{4\pi} {\cal A}^{\alpha+\frac{3}{2}} e^{-\beta\Phi} e^{-3\phi/2} 
\left(
e^\Phi-e^\phi \right)^{\alpha+\frac{3}{2}} \nonumber \\ & \hspace{3em} \times
{_2F_1}\left( \beta  , \alpha+1; \alpha+\frac{5}{2};\frac{e^\Phi-e^\phi}{e^\Phi}
\right)\frac{M}{a^3},
\end{align}
with $n$ the stellar proper number density. The proper mass of the
cluster is then \beqn M_p(r) = 4\pi \int_0^r \frac{n(r)m r^2 dr}{\sqrt{
    1-\frac{2GM}{c^2r}}}.  \neqn The difference between the total
proper mass $M_p(R)$ and the total gravitating mass $M(r)$ can be
interpreted as the gravitational binding energy of the cluster. We
will henceforth use the fractional binding energy \beqn f =
\frac{M_p(R)-M(R)}{M_p(R)} \neqn as a measure for the stability of a
cluster since analytical and numerical work has shown that radial
instability sets in in clusters around the first maximum of $f$
\citep{fa69,st85}.

From $p_\mu p^\mu = (mc)^2$, $J=r^2 p^3=mru^\phi$, and $p^0c = E e^{-\phi}$, it
follows that for a circular orbit in the $\theta=\pi/2$ plane the
angular momentum is given by \beqn J = \frac{r}{c} E \sqrt{ e^{-\phi}
  - \left( \frac{mc^2}{E} \right)^2 }. \neqn For a given radius $r$,
the energy of the circular orbit with that radius can be found by
setting $dJ/dr=0$. This leads to \beqn \frac{mc^2}{E} = \sqrt{
  \left(1-\frac{r}{2} \frac{d\phi}{dr} \right) e^{-\phi} }.  \neqn
Plugging this into the expression for the angular momentum yields
\beqn \frac{u^\phi}{c} = \frac{E}{mc^2} \sqrt{ \frac{r}{2}
  \frac{d\phi}{dr} e^{-\phi}}.  \neqn From the viewpoint of a distant observer,
the velocity of a star on a circular orbit with radius $r$ is given by
\beqn v_{\sf circ}(r) = \frac{d\tau_P}{dt} u^\phi = \frac{mc^2}{E} e^\phi
u^\phi = c\sqrt{ \frac{r}{2} \frac{d\phi}{dr} e^\phi} \label{vcgr} \neqn since the
derivative of the star's proper time $\tau_P$ with respect to
coordinate time is \beqn \frac{d\tau_P}{dt} = \frac{mc}{p^0} =
\frac{mc^2}{E} e^\phi.  \neqn

The radiation of a light source at rest at radius $r$, is observed at
infinity to have undergone a gravitational redshift \beqn z(r) =
e^{-\phi(r)/2}-1.  \neqn This ``redshift-from-rest'' is a measure for
how ``relativistic'' a given cluster model is.  Where it was first
  thought that no stable models with a central redshift-from-rest
  $z(0) \gtrsim 0.5$ can exist \citep{zp65,ip69,os74}, more recent
  work has shown that arbitrarily large values for the central
  ``redshift-from-rest'' are possible in stable models. The first hint
  that large redshifts are possible came from numerical integrations
  of the relativistic Boltzmann equation \citep{rst89} that was later
  on backed up by detailed analytical calculations
  \citep{mr95}. It was subsequently shown in \citet{bk98,bkm06} that arbitrarily
  large central redshifts are possible in stable models with a
  distribution function of the form $F(E) \propto \exp(-E/T)$, with
  $T$ the uniform kinetic temperature as observed from infinity, only
  if $T/mc^2 \lesssim 0.06$. ``Hotter'' models are stable only for
  red\-shifts below $\approx 0.5$.  

\subsection{The Newtonian limit}

In the Newtonian limit, we can employ the approximation
\begin{align}
\frac{d\tau_P}{dt} &= \frac{mc^2}{cp^0} = \frac{mc^2}{E}e^\phi
\nonumber \\ &= \frac{1}{c}\frac{ds}{dt}\approx \sqrt{ e^\phi - \left(
  \frac{v}{c} \right)^2},
\end{align}
or, in other words, \begin{align} E &\approx \frac{mc^2 e^\phi}{\sqrt{ e^\phi -
    \left( \frac{v}{c} \right)^2}} \approx mc^2\left( 1 +
\frac{1}{2}\phi + \frac{1}{2} \frac{v^2}{c^2} \right) \nonumber \\ &\approx mc^2 +
m\left( \psi + \frac{1}{2}v^2 \right) = mc^2 + m \epsilon.
\end{align}
Here, $\epsilon$ is the Newtonian energy per unit mass. Moreover,
$\hat{p}_0c \approx mc^2+\frac{1}{2}mv^2$. 

These results can be used to calculate the Newtonian approximation for
the isotropic momentum moments of the DF, given by expression (\ref{mu2q}):
\begin{align} \mu_{k,2q} &\approx 4 \pi (mc)^{k-1} \int F(E) p^{2q+2} dp \nonumber \\
&\approx (mc)^{k-1} \mu^N_{2q}(\psi). \label{mckf}
\end{align} Except for the inconsequential forefactor $(mc)^{k-1}$,
  this is the correct expression for the Newtonian isotropic momentum
  moment $\mu^N_{2q}(\psi)$.

Taking together $E \approx E_0 \approx mc^2$, $D_{E^2} \approx
\frac{1}{2m^2c^2} D_\epsilon$, $dE_0^2 \approx 2 m^2 c^2 d\psi$, and
(\ref{mckf}), the inversion formula for the DF can be written in the
form \beqn F(\epsilon) \approx
\frac{1}{2^{q+2}\pi^{\frac{3}{2}}\Gamma\left(q+\frac{3}{2}\right)
  m^{2q+3}} D_\epsilon^{q+2} \int
\frac{\mu^N_{2q}(\psi)}{\sqrt{2(\psi-\epsilon)}} d\psi, \neqn the
correct Newtonian expression for the DF in terms of a Newtonian
momentum moment. For $q=0$, one obtains the important special case
\beqn F(\epsilon) \approx \frac{1}{2\pi^2m^3} D_\epsilon^2 \int
\frac{\rho(\psi)}{\sqrt{2(\psi-\epsilon)}} d\psi, \neqn with $\rho$
the mass density.

In the Newtonian limit, the distribution function (\ref{propdf}) becomes $F(E)
\approx f_0 \left[ 2(\Psi-\epsilon) \right]^\alpha$, with
$\Psi=\psi(R)$, the value of the Newtonian gravitational potential at
the outer edge of the cluster.  For an infinitely extended system with
$\alpha=7/2$ and $\Psi=\psi(\infty)=0$ this is $f_{\text{Plum}}(E)=f_0
(-2\epsilon)^{7/2}$, the distribution function of the Newtonian
Plummer model. The Newtonian limit of the distribution function does
not depend on the parameter $\beta$:~it only serves to change the
slope of the DF for the most strongly relativistic models. In the
Newtonian limit, the density reduces to \beqn \rho(\psi) \approx
\frac{3}{4\pi} \left( -\frac{a}{GM} ( \Psi-\psi )
\right)^{\alpha+\frac{3}{2}}\frac{M}{a^3}. \neqn For a Plummer model,
with $\alpha=7/2$, we retrieve the relation \beqn \rho_{\text{Plum}}
\approx \frac{3}{4\pi} \left( -\frac{a}{GM} \psi
\right)^5\frac{M}{a^3}.  \neqn The proper density $nm$ reduces to the
same expression as the gravitating mass density $\rho$, as it
should. The Newtonian expression for the pressure is found to be \beqn
P \approx \frac{3}{2\pi(2\alpha+5)} \left( -\frac{a}{GM} (\Psi-\psi)
\right)^{\alpha+\frac{5}{2}}\frac{GM^2}{a^4}.  \neqn For a Plummer
model, we find \beqn P_{\text{Plum}} \approx \frac{1}{8\pi} \left(
-\frac{a}{GM} \psi \right)^6\frac{GM^2}{a^4} \propto
\rho_{\text{Plum}}^{\frac{6}{5}}.  \neqn

Clearly, these Newtonian models have equations of state of the form
\beqn P = K \rho^{\frac{2\alpha+5}{2\alpha+3}} = K
\rho^{1+\frac{1}{n}} \neqn for some constant $K$. They are polytropes
with polytropic index \beqn n = \alpha+\frac{3}{2} \ge 0. \label{nalpha} \neqn The
general relativistic cluster models, due to the presence of the
hypergeometric functions in the expressions for the density and
pressure, are not polytropes and have more complicated equations of
state. Newtonian polytropes have finite mass for $n \in [0,5]$ and
finite radius for $n \in [0,5[$. The Plummer model, with $n=5$ is the
    first polytropic model with infinite radius but still with finite
    mass.  It is generally assumed that the condition $df/dE<0$ is a
    prerequisite for the radial stability of a cluster
    \citep{ip69,fa71}. We therefore limit ourselves to models with
    $\alpha \ge 0$, and hence $n \ge \frac{3}{2}$, for which this
    condition is definitely fulfilled.

As is well known, the structure of a polytrope with index $n$ and
equation of state $P = K \rho^{1+\frac{1}{n}}$ for some constant
forefactor $K$ is given by the Lane-Emden equation \beqn
\frac{1}{\xi^2} \frac{d}{d\xi} \left( \xi^2 \frac{d\theta}{d\xi}
\right) = - \theta^n. \neqn Here, $\xi$ is a dimensionless radius
related to the radius $r$ via \beqn \xi = \sqrt{ \frac{2 \pi c^2
    a^3}{(n+1){\cal A}M}\frac{\rho_c^2}{P_c} } \frac{r}{a} \label{xix}, \neqn
with $\rho_c$ and $P_c$ the central density and pressure,
respectively. This equation must be integrated numerically for the
function $\theta(r)$ out to its first zero, which then defines the
outer radius $R$ of the cluster. Then the density is given by $\rho(r)
= \rho_c \theta^n(r)$, and the gravitational potential by $\psi(r) =
-(n+1)K \rho_c^{\frac{1}{n}} \theta(r) + \psi(R)$. The circular
velocity profile, $v_{\sf circ}(r)$, then follows from the relation
\beqn v_{\sf circ}(r) = \sqrt{ r \frac{d\psi}{dr} }.  \neqn

\begin{figure}
\includegraphics[width=0.49333\textwidth]{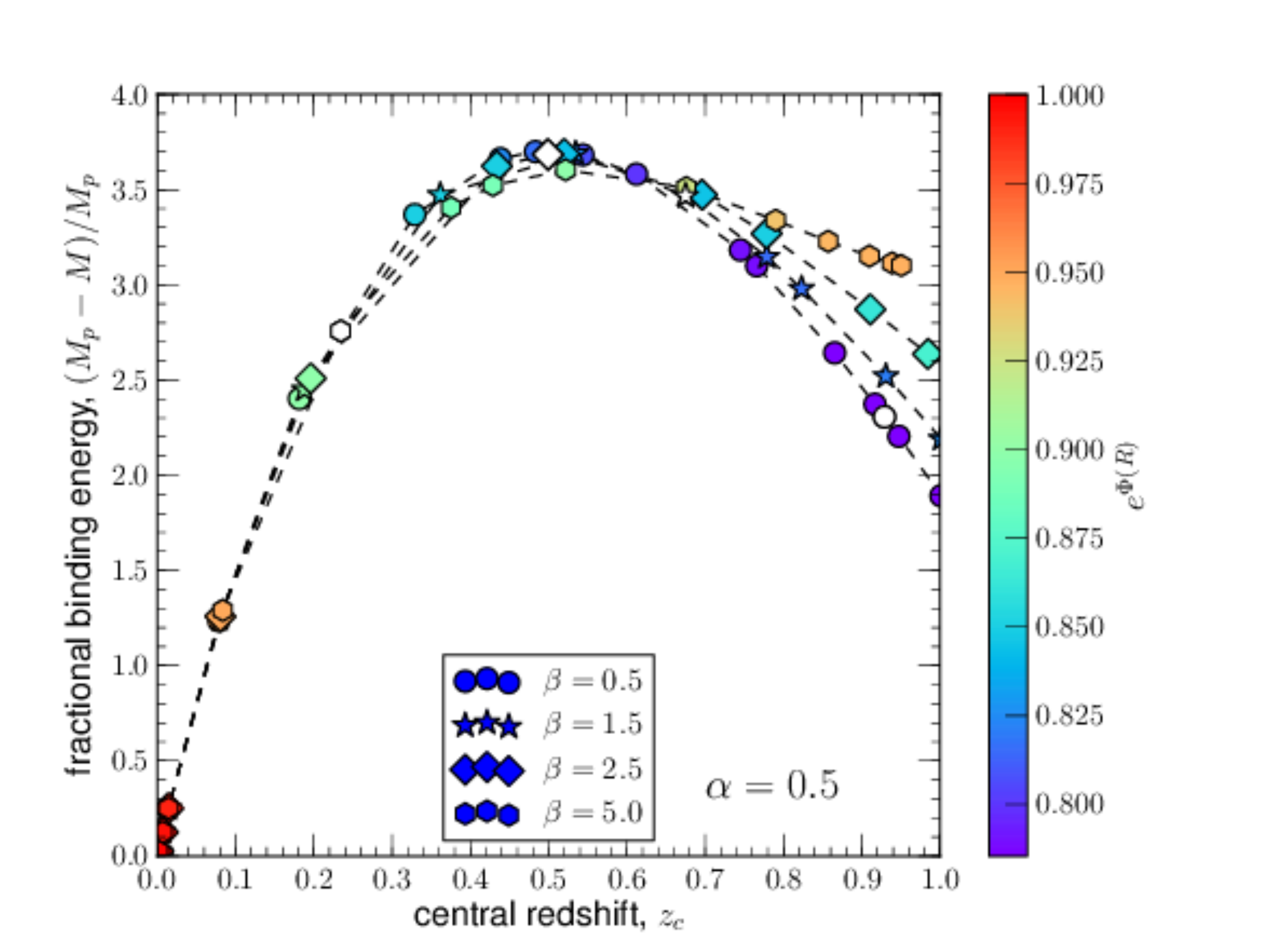}
\includegraphics[width=0.49333\textwidth]{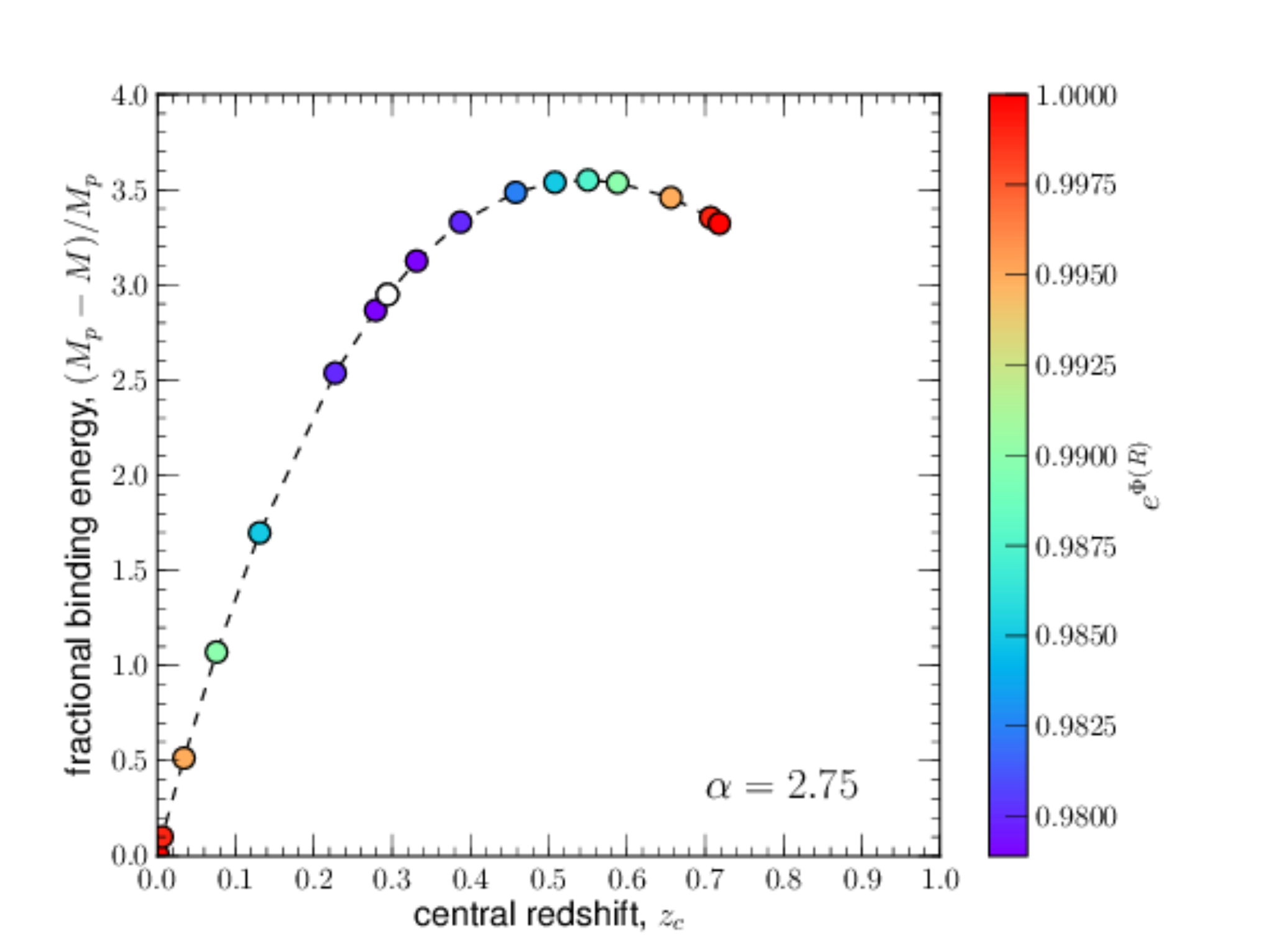}
\includegraphics[width=0.49333\textwidth]{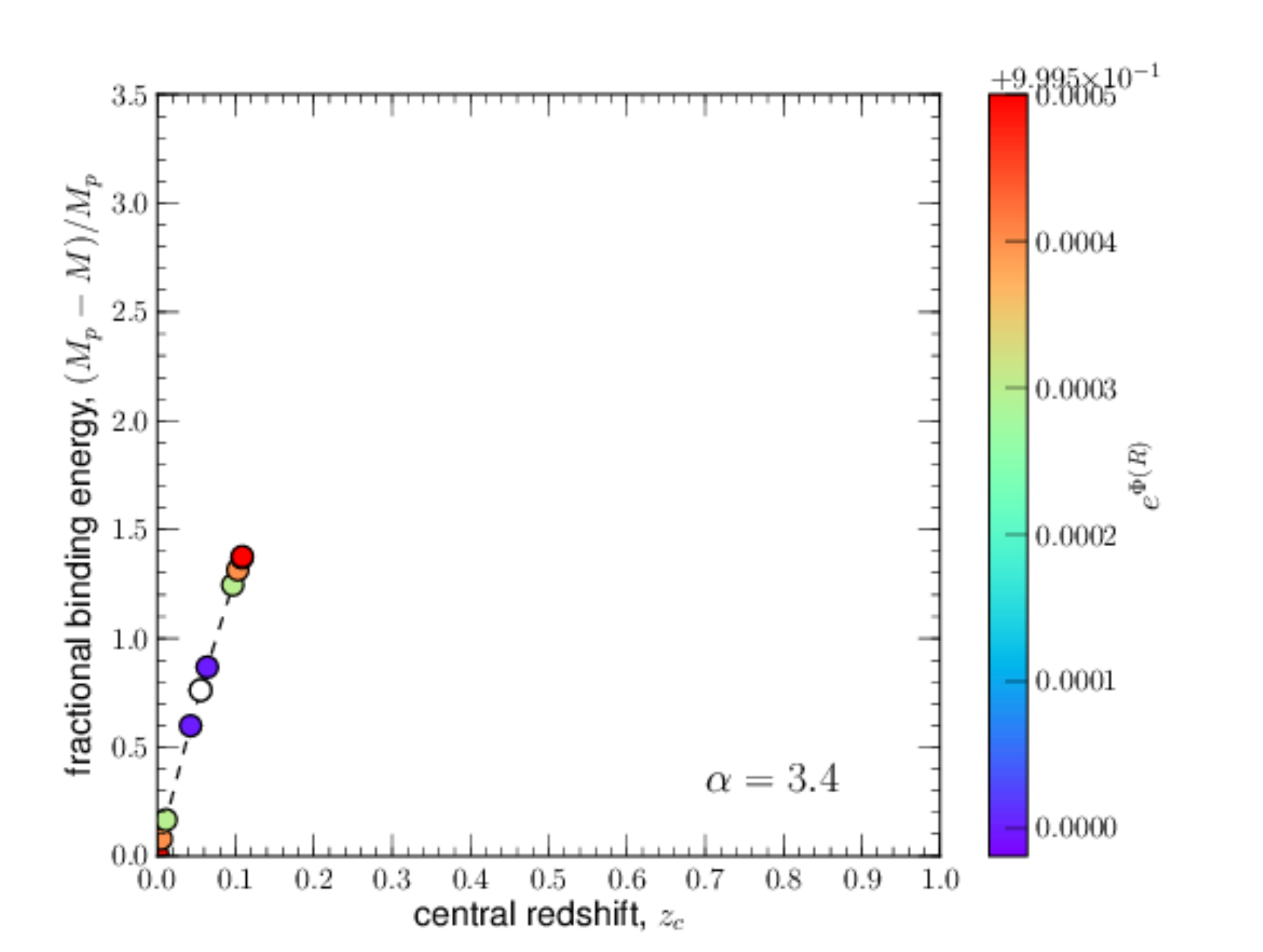}
\caption{Central redshift-from-rest versus fractional binding energy
  $f=(M_p(R)-M(R))/M_p(R)$ for all models with $\alpha=0.5$ (top),
  $\alpha=2.75$ (middle), and $\alpha=3.4$ (bottom). For $\alpha=0.5$,
  the effect of different $\beta$-values, between 0.5 and 5.0, is
  explored. For all other $\alpha$-values, only $\beta=0.5$ was used.
  The color scale of the data points indicates the value of the
  potential at the outer boundary of the model, $e^\Phi$. The model
  with the smallest value for $e^\Phi$ for each $\alpha$-value is
  indicated with a white dot in each panel. Models to the left of this
  white dot have shallower potentials; those to the right of it have
  deeper potentials.
 \label{fig:bifurc_1.67.eps}}
\end{figure}
With which Newtonian model should a given relativistic cluster be
compared? A natural choice for the polytropic index is given by
(\ref{nalpha}). From (\ref{xix}), it is obvious that the dimensionless
radius $\xi$ can be rescaled to the dimensionless radius $x$ with the
scale depending on the central pressure and density. We rescale the
density profile such that the total mass equals unity, something we
will also do with the relativistic models, giving us a value for
$\rho_c$. We then adopt a value for the constant $K$ such that $P_c =
K \rho_c^{1+\frac{1}{n}}$. In this case, $2\psi(R)/c^2 = -1/{\cal A}X$
with $X=R/a$ the dimensionless outer boundary of the Newtonian cluster
(which, obviously, does not need to coincide with the outer boundary
of the relativistic cluster).

One further remark concerns the fact that in the case of Newtonian
stellar clusters, one can choose the mass-scale $M$ and the
length-scale $a$ independently from each other whereas in the general
relativistic models presented here these two parameters are linked by
the parameter $\cal A$, defined as (\ref{Qparam}), and they cannot be
chosen freely. However, in the Newtonian limit, which can be defined
formally as the limit $c \rightarrow \infty$, the parameter $1/{\cal A}$
goes to zero, \beqn \lim_{c \rightarrow \infty} \frac{1}{\cal A} = \lim_{c
  \rightarrow \infty} \frac{2GM}{c^2a} = 0, \neqn for a finite
mass-scale $M$ and non-zero length-scale $a$. In that limit, $M$ and
$a$ are effectively decoupled since $1/\cal A$ is always zero, irrespective of
which mass and length-scale one chooses.

\section{Solving the field equations} \label{solve}

The two relevant field equations, as shown in e.g. \cite{os74}, are
\begin{align}
\frac{dM}{dr}(r) &= 4 \pi r^2
\rho \label{cofi00a}\\ \frac{d\phi}{dr}(r) &= \frac{2G}{c^2r^2}
\left[M(r)+ 4\pi r^3
    \frac{P}{c^2}\right]
\left[1-\frac{2GM(r)}{rc^2}\right]^{-1} . \label{cofi11a}
\end{align}
If we denote the dimensionless radius by $x=r/a$, the scaled mass by
${\cal M}(r)=M(r)/M$, the scaled density by $\tilde{\rho} = \rho
a^3/M$, and the scaled pressure by $\tilde{P} = P a^3/Mc^2$, we can
rewrite these equations in a fully dimensionless form as
\begin{align}
\frac{d{\cal M}}{dx} &= 4 \pi x^2 \tilde{\rho} \\
\frac{d\phi}{dx} &= \frac{1}{{\cal A}x^2} \left[ {\cal M} +
4\pi x^3 \tilde{P} \right]\left[ 1 - \frac{\cal M}{{\cal A}x} \right]^{-1}.
\end{align}
These equations must be integrated numerically starting from the
initial conditions
\begin{align}
{\cal M}(0) &= 0, \nonumber \\
\phi(0) &= \phi_0
\end{align}
where $\phi_0$ must be chosen such that \beqn \exp(\phi(X)) =
\exp(\Phi) = 1-\frac{{\cal M}(X)}{{\cal A}X}, \label{ephiqm} \neqn with $X=R/a$ the scaled
radius at the cluster's outer boundary. This ensures that the
``internal'' solution smoothly goes over into the ``external''
Schwarzschild solution. This precludes the retrieval of infinitely
extended models, especially if they have a diverging mass.

\begin{figure*}
\centering
\includegraphics[width=0.85\textwidth]{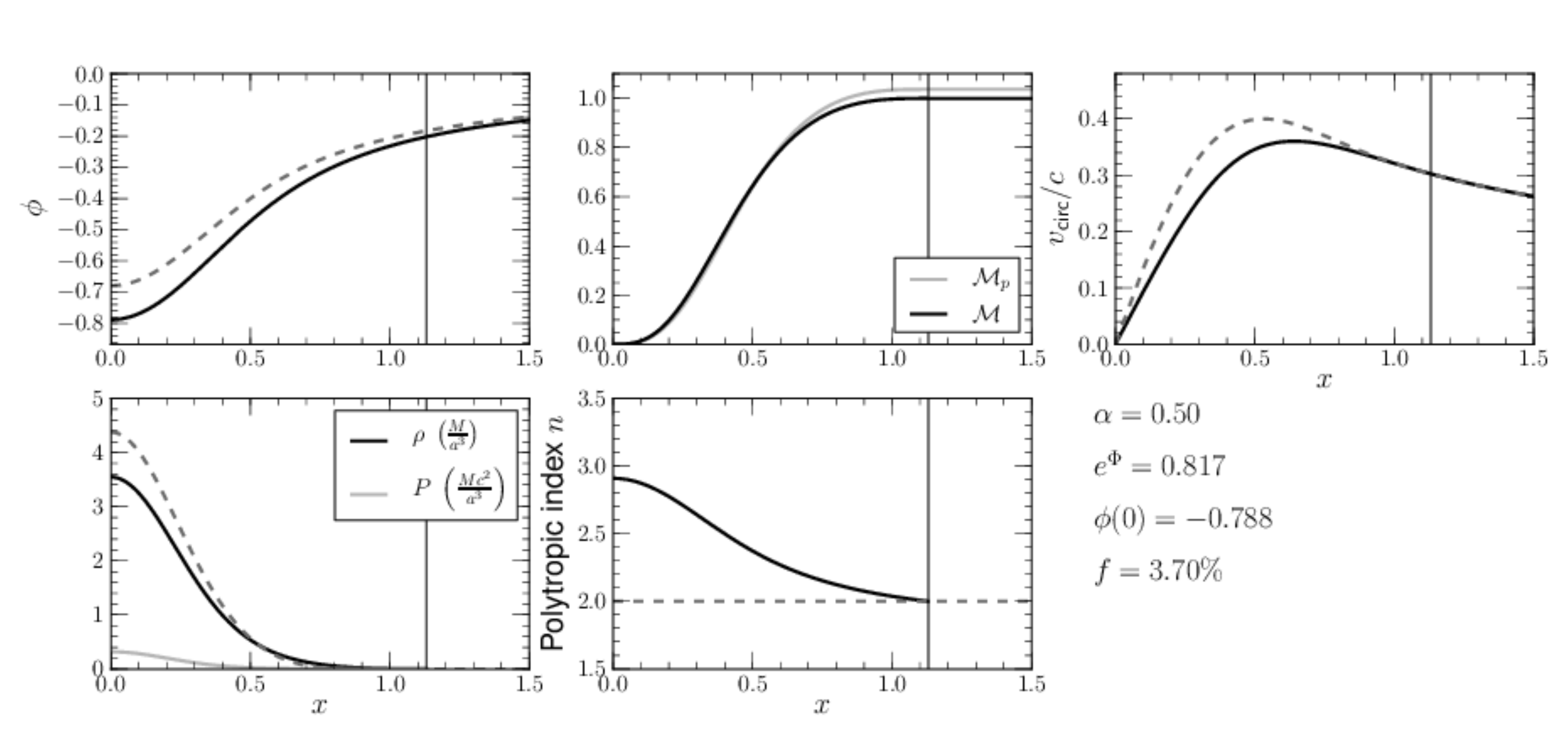}
\includegraphics[width=0.85\textwidth]{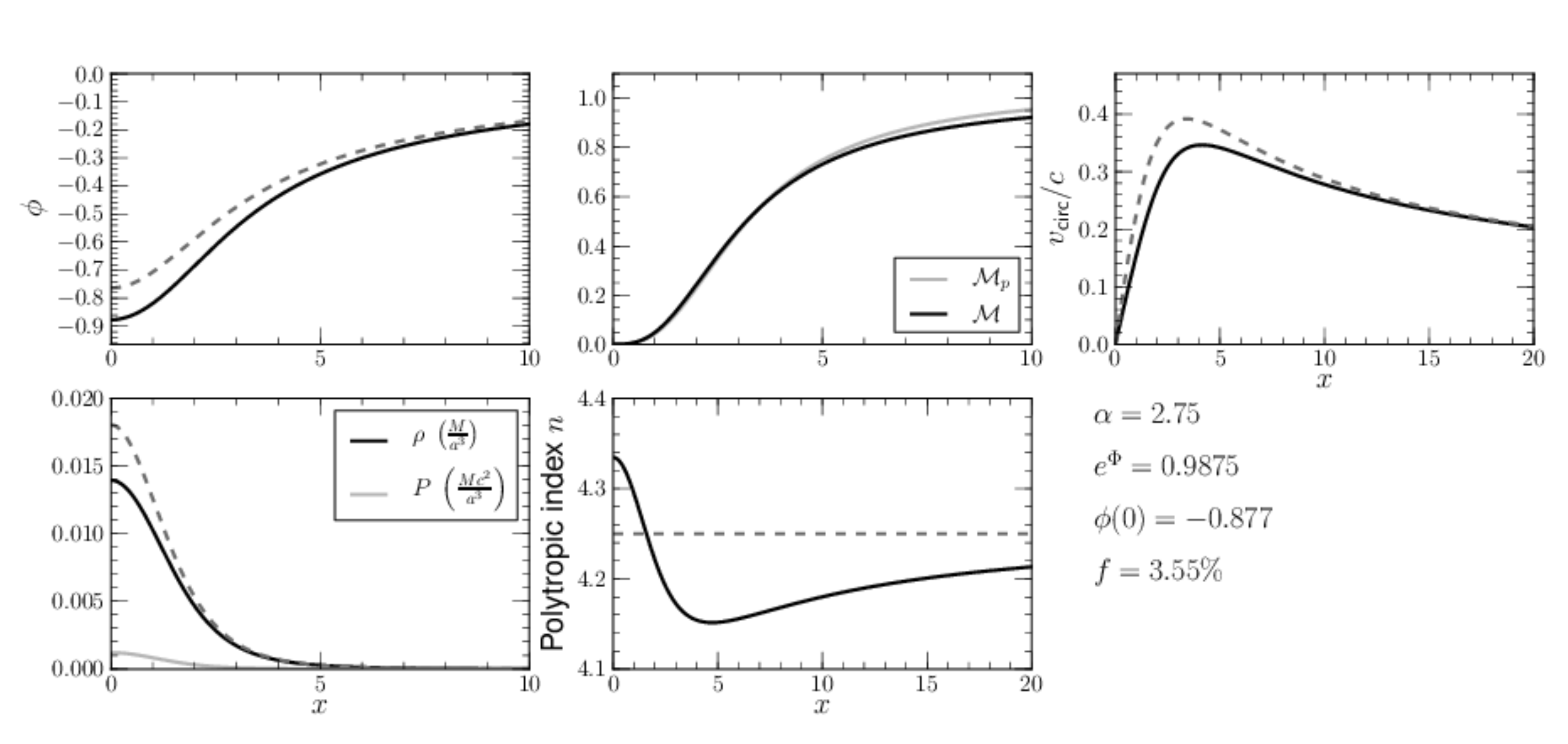}
\includegraphics[width=0.85\textwidth]{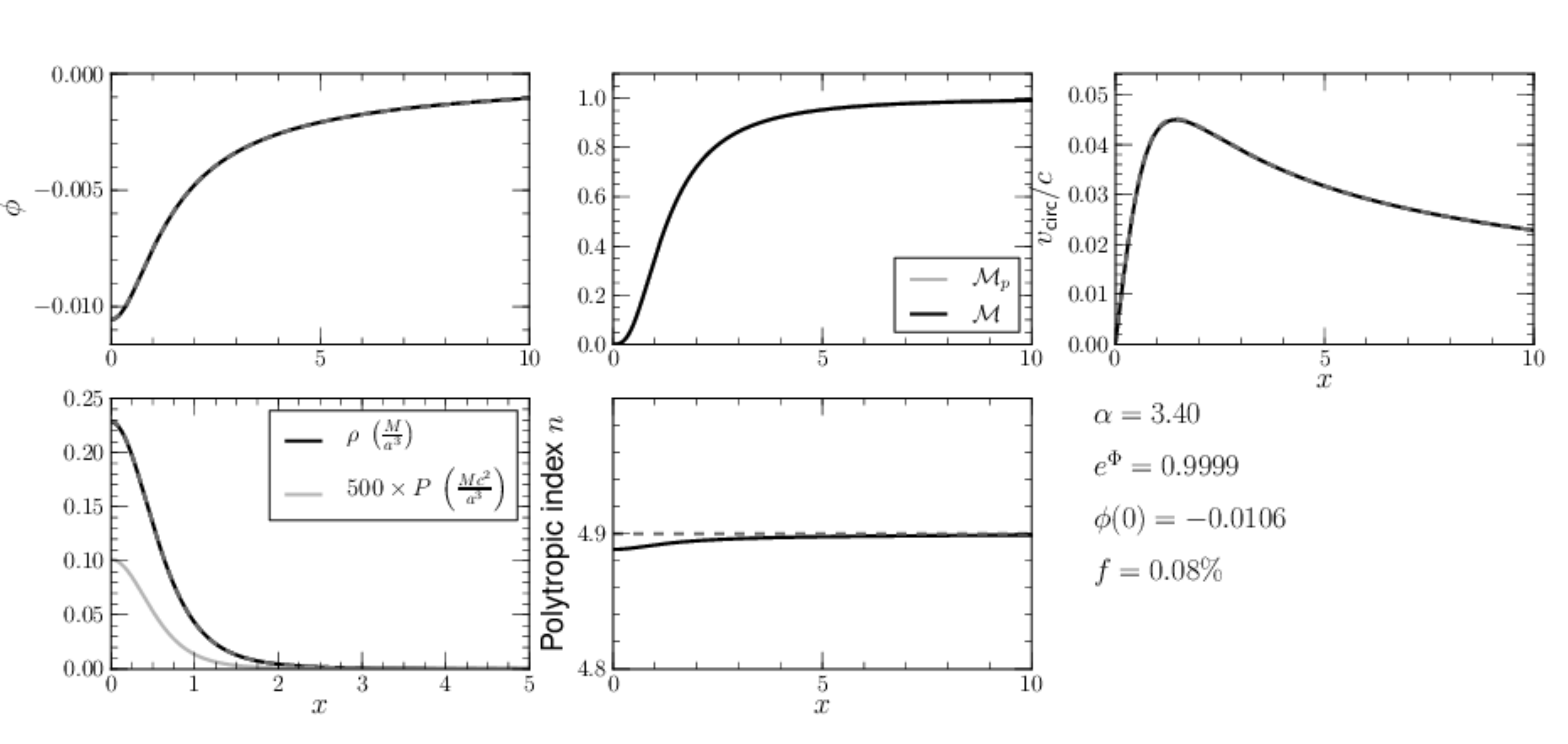}
\caption{The potential, $\phi$, gravitating and proper mass profiles,
  ${\cal M}$ and ${\cal M}_p$, circular-velocity profile, $v_{\sf
    circ}$, density $\rho$ and pressure $P$ profiles, and effective
  polytropic index $n$ for models with $\alpha=0.5$, 2.75, and
  3.4. Where visible, the vertical line indicates the outer boundary
  of the cluster model. The dashed curves indicate the
  circular-velocity profile, density profile, and polytropic index of
  the corresponding Newtonian cluster with the same total gravitating
  mass and the same central pressure. For each model, its values for
  $\alpha$, the boundary potential $e^\Phi$, the central potential
  $\phi(0)$, and the fractional binding energy $f$ are indicated. All
  models have $\beta=1/2$.
 \label{fig:models.eps}}
\end{figure*}
By explicitly pulling out the $\cal A$-dependence of the density and
pressure, it becomes clear that by rescaling the mass and radius
according to
\begin{align}
x' &= {\cal A}^{(1+2\alpha)/4}x \\
{\cal M}' &= {\cal A}^{(2\alpha-3)/4}{\cal M},
\end{align}
the parameter $\cal A$ can be completely removed from the
dimensionless field equations. So one can always set ${\cal A}=1$ in
the field equations, solve them, and then afterwards rescale to that
particular value of ${\cal A}$ for which ${\cal M}(X)=1$. In that
case, the mass scale $M$ equals the total gravitating mass of the
cluster and ${\cal A}$ has the meaning of the ratio of $a$ to $R_S$.

We wrote a small Python program to numerically integrate these
equations and to determine the central value of the potential using a
least squares minimizer.

\section{Discussion} \label{discussion}

\subsection{Existence of solutions} \label{sub:exist}

For each choice of $\alpha$, the only free parameter in the field
equations is the value of the potential at the outer boundary of the
cluster, in the form $e^\Phi$. Our numerical work shows that the field
equations presented in the previous section have a bifurcation at some
$\alpha$-dependent critical value, $e^{\Phi_0(\alpha)}$. For $e^\Phi <
e^{\Phi_0(\alpha)}$, no solutions exist. At $e^\Phi =
e^{\Phi_0(\alpha)}$, a single solution appears. For choices $1 \ge
e^\Phi > e^{\Phi_0(\alpha)}$, two solutions, with different central
potential values $\phi_0$, exist. This can be seen in
Fig. \ref{fig:bifurc_1.67.eps} in which the fractional binding energy
$f$ is plotted versus the central redshift-from-rest $z_c$ for models
with $\alpha=0.5$, 2.75, and 3.4. We always adopt the value
$\beta=1/2$ except in the top panel, where the effects of different
$\beta$-values are explored. In each panel, the model with the
smallest value for $e^\Phi$ is indicated by a white dot. The color of
the other data points corresponds to their $e^\Phi$-value, as
indicated by the colorbar. To the left of the white dot are models
with shallower potential wells with the Newtonian $f=0$, $z_c=0$ model
as limit. To the right of the white dot are models with deeper
potential wells and correspondingly higher central redshifts. 
  This situation is reminiscent to that of the family of models
  discussed by \citet{bk98} which also exhibits both bifurcations
  (i.e. more than one solution for a given set of model parameters)
  and limiting values for a parameter connected to the energy at the
  outer boundary. 

For small values for the power $\alpha$, below $\alpha \sim 3$, the
$f-z_c$-curve has a maximum around $z_c \approx 0.5$. As $\alpha$
increases, the right side of the $f-z_c$-curve appears to curl up from
right to left until this maximum disappears and the $f-z_c$-relation
is monotonically rising. In the limit $\alpha \rightarrow 7/2$, only
the Newtonian $f=0$, $z_c=0$ model exists. This means that the Plummer
model is a purely Newtonian construct:~no relativistic models with
$\alpha=7/2$ exist. Also in the Newtonian is the Plummer model a
limiting case. As the polytropic index $n$ is increased from zero, it
is the first solution of the Lane-Emden equation with infinite
extent. It is also the last model with a finite total mass. This
appears also to be true relativistically. By construction we are
searching for models with a finite total mass by trying to match the
solutions of the field equations to an external Schwarzschild
metric. No such solutions exist for $\alpha > 7/2$.

This is true for different values of the power $\beta$. However,
increasing $\beta$ shifts the high-$z_c$ end of the $f-z_c$-relation
in the direction of smaller $z_c$, i.e. towards models with shallower
potentials. An increase of $\beta$ also raises the $e^\Phi$-value of
those most relativistic cluster models which means they become less
compact (see paragraph \ref{sub:radius}). 

The general conclusion we can draw from this is that the steeper the
distribution function $F(E)$ varies as a function of energy $E$, the
more the solutions are confined towards the Newtonian limit
$(f=0,z_c=0)$ and that no relativistic models with finite mass exist
with $\alpha>7/2$.

\subsection{Model properties}

\begin{figure*}
\includegraphics[width=\textwidth]{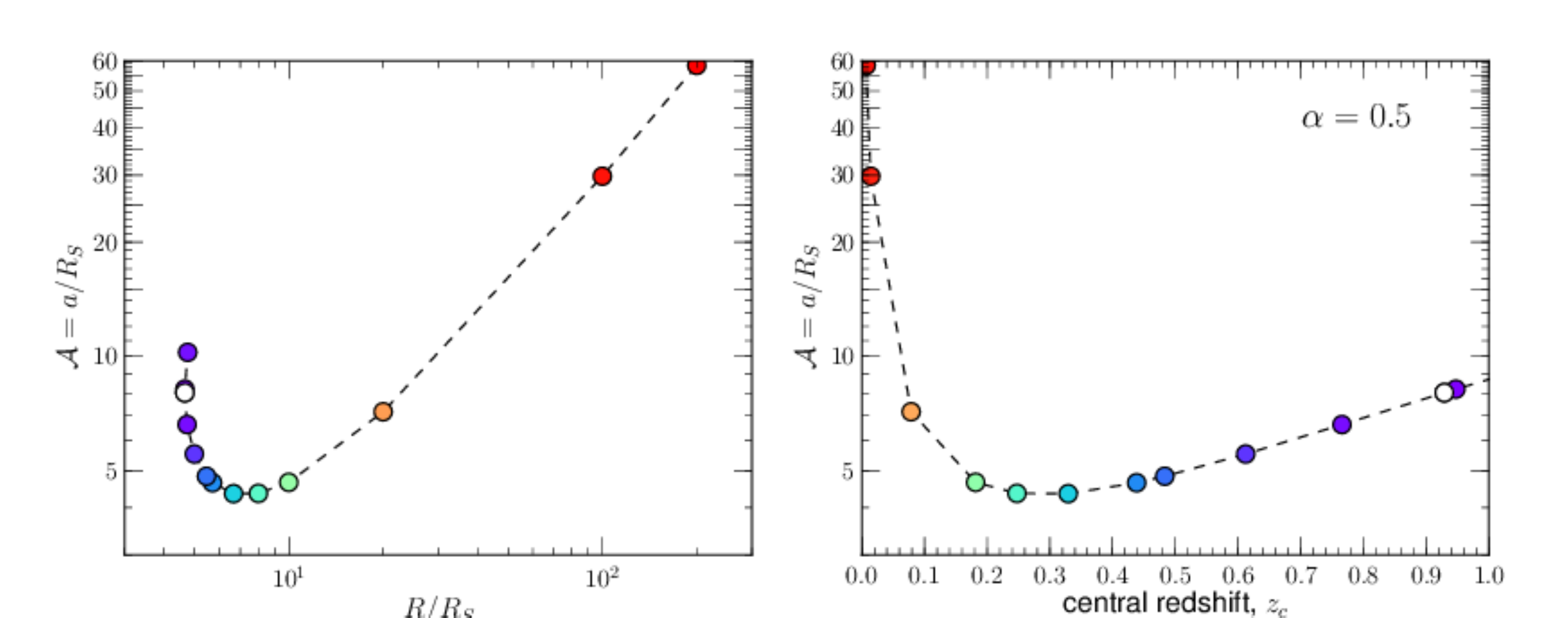}
\caption{The ratio of the scale radius to the Schwarzschild radius,
  ${\cal A}=a/R_S$ versus outer boundary radius $R/R_S$ (left panel)
  and the central redshift-from-rest $z_c$ (right panel) for the
  models with $\alpha=\beta=0.5$. The color scale of the data points
  indicates the value of the potential at the outer boundary of the
  model, $e^{\Phi}$.
 \label{fig:A_zc.eps}}
\end{figure*}

In Fig. \ref{fig:models.eps}, we present the potential function
$\phi$, the gravitating and proper mass profiles, ${\cal M}$ and
${\cal M}_p$, the circular-velocity profile, $v_{\sf circ}$, the
density $\rho$ and pressure $P$ profiles, and the effective polytropic
index $n$ for models with $\alpha=0.5$, 2.75, and 3.4. For all models,
we adopt $\beta=1/2$. The dashed curves indicate the circular-velocity
profile, density profile, and polytropic index of the corresponding
Newtonian cluster with the same total gravitating mass and the same
central pressure. The effective polytropic index $n$ is here defined
as \beqn 1+\frac{1}{n} = \frac{d\ln P}{d\ln \rho} \neqn which can be
compared with the index (\ref{nalpha}) derived from the power $\alpha$
in the expression for the distribution function. In the Newtonian
limit, both indices coincide.

The $\alpha=3.4$ model shown in Fig. \ref{fig:models.eps} has a very
shallow potential and is essentially Newtonian. Therefore, it is
indistinguishable from the Newtonian solution of the Lane-Emden
equation. The models with $\alpha=0.5$ and $\alpha=2.75$ have much
deeper gravitational wells and are well in the general relativistic
regime. Clearly, these models do not have polytropic equations of
state and their effective polytropic indices can differ significantly
from the value expected from their $\alpha$-value. For the same total
mass, their density profiles are less steep than those of the
Newtonian models. This, combined with the gravitational time
dilatation effect in eqn. (\ref{vcgr}) for the circular velocity,
causes the relativistic circular-velocity curve to be much flatter
than its Newtonian counterpart.

\begin{figure}
\includegraphics[width=0.53\textwidth]{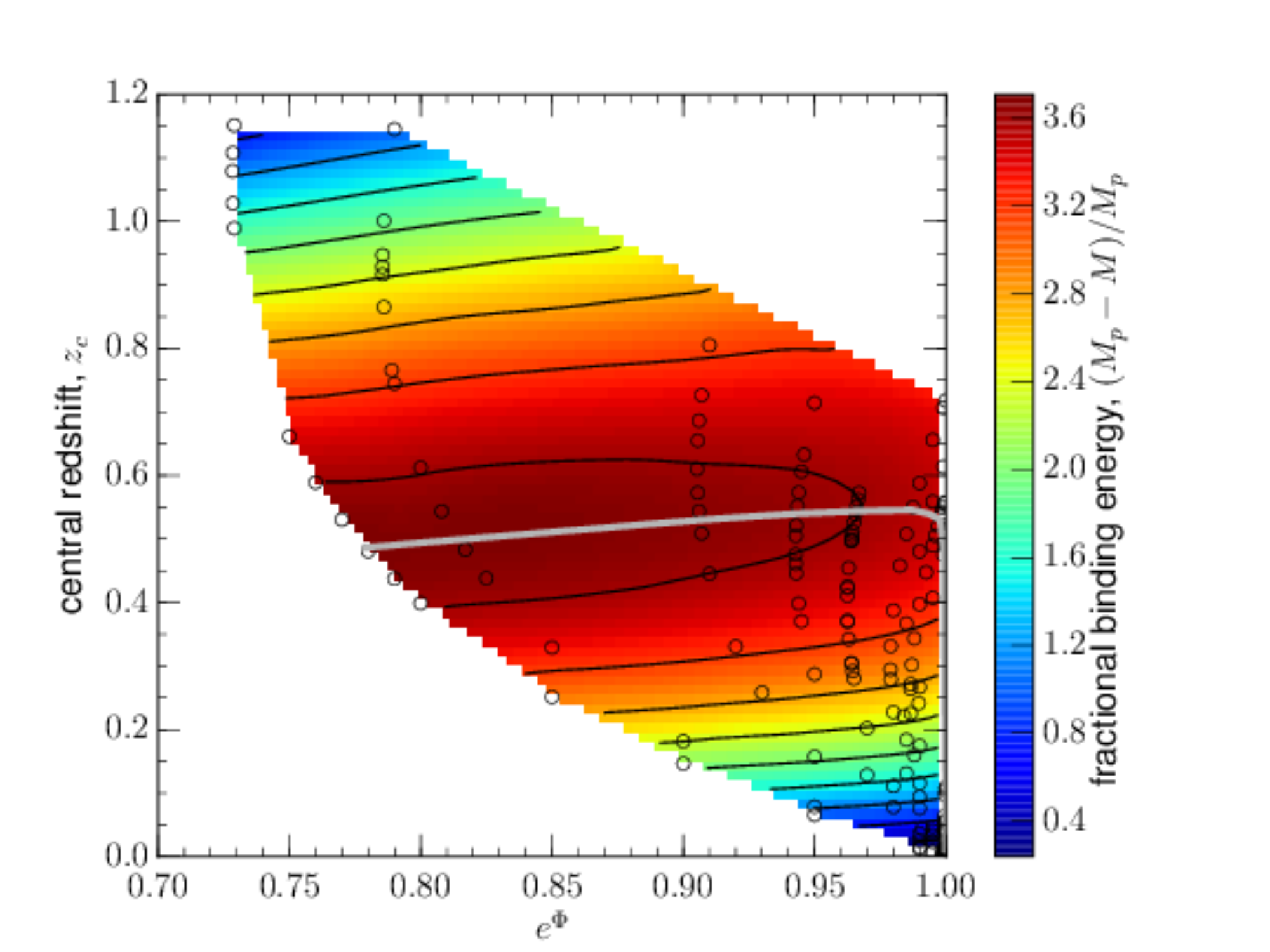}
\caption{The fractional binding energy $f = (M_p(R)-M(R))/M_p(R)$ as a
  function of central redshift $z_c$ and the potential at the outer
  edge of the mass distribution, quantified by $\exp(\Phi)$. The color
  scale measures $f$ in percentages; the open circles indicate the
  positions of the models that were actually constructed. The
  different model sequences have different values for the power
  $\alpha$.
 \label{fig:z_f.eps}}
\end{figure}

The ratio of the scale-length to the Schwarzschild radius, quantified
by ${\cal A}=a/R_S$, is plotted as a function of the ratio of the
outer boundary radius to the Schwarzschild radius, $R/R_S$, and of the
central redshift-from-rest, $z_c$, in Fig. \ref{fig:A_zc.eps} for the
$\alpha=\beta=0.5$ models. $a/R_S$ shows a non-trivial behavior in the
sense that the model with the smallest scale-length is neither the
most tightly bound model (the one with the largest fractional binding
energy $f$) nor the most compact one (the one with the smallest
$e^\Phi$ or $R/R_S$ value). $a/R_S$ diverges for $z_c \rightarrow 0$
since the Schwarzschild radius tends to zero in the Newtonian
limit. In the limit of extremely compact models, $a/R_S$ increases
again. Apparently, only models with very flat-topped density profiles,
with $a \gtrsim R$, can exist in this regime.

\subsection{Binding energy}

We plot the fractional binding energy $f = (M_p(R)-M(R))/M_p(R)$ as a
function of central redshift $z_c$ and the potential at the outer edge
of the mass distribution, quantified by $\exp(\Phi)$, in
Fig. \ref{fig:z_f.eps}. The open circles in this figure indicate the
loci of the models that were actually constructed. The different model
sequences have different values for the power $\alpha$, the leftmost
corresponding to $\alpha=0.05$. The 2D map of the binding energy was
constructed by applying a bicubic spline interpolator to the model
points. The models nicely cover the first maximum of $f$, where
dynamical instability is expected to set in \citep{bk98,bkm06}.

The grey line connects the models which, for a given $\alpha$, attain
the maximum fractional binding energy. The models with $\alpha$ in the
range $0.05$ to $\approx 3.0$ have central redshift-from-rest values
between $\approx 0.5$ and $\approx 0.55$. For higher $\alpha$-values,
the maximum central redshift rapidly drops to zero. As the power
$\alpha$ approaches the value of 7/2, the Plummer model value, both
the central redshift-from-rest and the fractional binding energy go to
zero, the Newtonian limit. The overall maximum central
redshift-from-rest is achieved by the model with $\alpha=2.75$. 

This behavior is caused by the $\alpha$-dependence of the shape of the
$f-z_c$-relation which was discussed in paragraph \ref{sub:exist}. At
first, steepening the distribution function by increasing $\alpha$
above zero leads to a deepening of the potential well and therefore to
a slight increase of $z_c$. Above $\alpha \approx 2.75$, a further
steepening of the distribution function and of the density profile
limits the models more and more to the Newtonian limit, thus reducing
$z_c$.

\subsection{The radius} \label{sub:radius}

Each model is labelled by a unique $\cal A$-value for which the mass
scale $M$ coincides with the model's total mass. If we select this
particular value for $\cal A$ or, equivalently, $M$ , the quantity
\beqn R_S = \frac{2GM}{c^2} \neqn has the physical meaning of being
the model's Schwarzschild radius. Numerically integrating the field
equations yields the dimensionless outer radius $X$. Multiplying this
radius with the scale-length $a$ gives the physical value for the
radius $R=aX$. It then follows that \beqn \frac{1}{\cal A} = \frac{2GM}{c^2a}
= \frac{2GM}{c^2R}X \neqn and consequently \beqn \frac{R}{R_S} =
{\cal A}X  = \frac{1}{1-\exp(\Phi)}, \neqn where we made
use of eqn. (\ref{ephiqm}).

\begin{figure*}
\includegraphics[width=\textwidth]{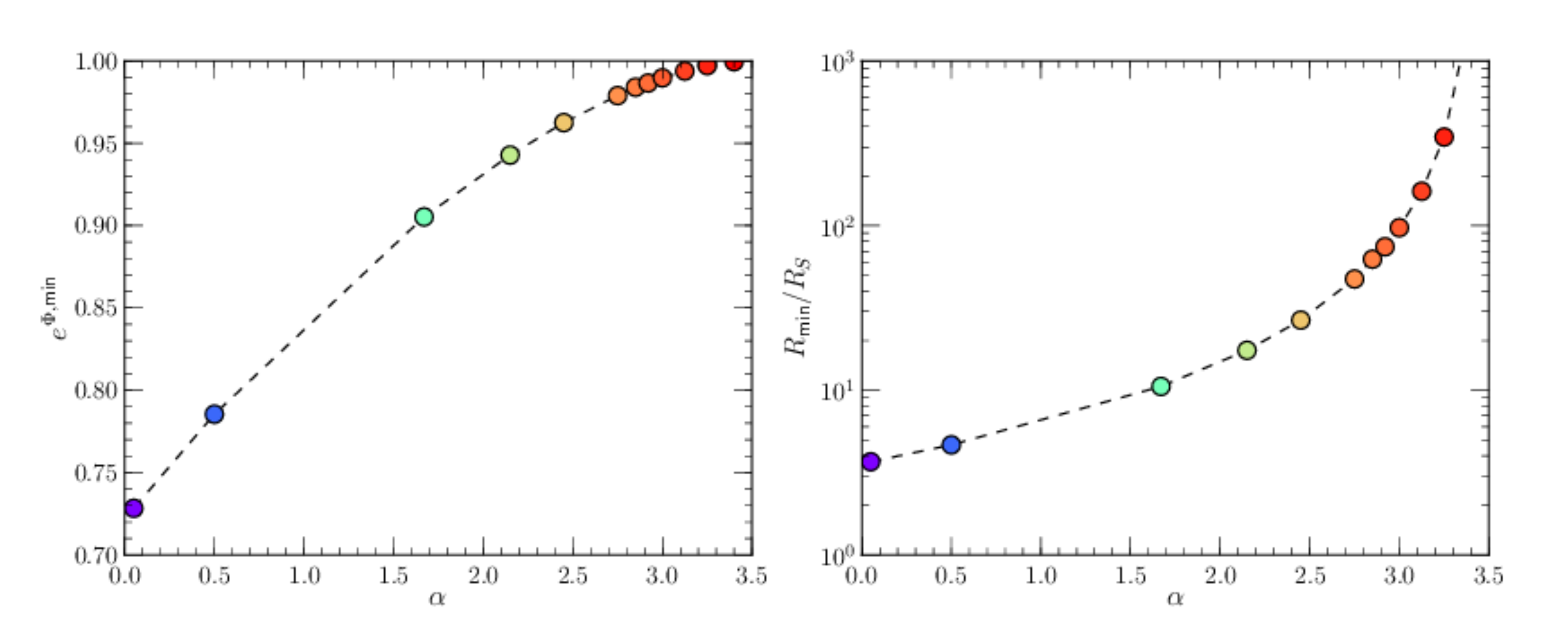}
\caption{The minimum value for $e^\Phi$ versus $\alpha$ (left) and the
  smallest possible radius, divided by the model's Schwarzschild
  radius versus $\alpha$ (right). 
 \label{fig:alpha_phimin_Rmin.eps}}
\end{figure*}

For each value of $\alpha$, there exists a minimum value for $\Phi$
below which no solutions to the field equations can be found. Using
the above, this corresponds to a minimum value for $R/R_S$. As
$\alpha$ tends to zero, the minimum radius shrinks to $R \approx 3.6
R_S$, as can be seen in Fig. \ref{fig:alpha_phimin_Rmin.eps}. Hence,
models with very ``flat'' distribution functions and density and
pressure profiles can be very small, with radii only a few times
larger than their Schwarzschild radius. As the distribution function
and the corresponding density and pressure profiles are steepened by
increasing the value of the power $\alpha$, this minimum radius
steadily increases. In the limit $\alpha \rightarrow 7/2$ the only
possible solution is the Newtonian Plummer model and the minimum
radius grows to infinity.

\section{Conclusions} \label{conc}

We show that the equations underlying the general relativistic theory
of spherically symmetric isotropic stellar clusters can be cast in a
form analogous to that of the Newtonian theory. Using the mathematical
formalism developed for the latter, we prove that the distribution
function can be derived from any isotropic momentum moment
$\tilde{\mu}_{k,2q}$. This is a direct generalization of the inversion
relations derived by \citet{fa68} and \citet{pk96}. Moreover, every
higher-order moment $\tilde{\mu}_{k,2q}$, with $q>0$, can be written
as an integral over the corresponding zeroth-order moment
$\tilde{\mu}_{k,0}$.

We propose a mathematically simple expression for the distribution
function of a family of isotropic cluster models which is guaranteed
to be positive everywhere in phase space. The distribution function of
each model is basically defined by two parameters:~the slope $\alpha$
and the value of the potential at the boundary, $\Phi$. In the
Newtonian limit, these models reduce to the family of polytropic
models. In the relativistic regime, however, these models do not have
a polytropic equation of state. We derive the Newtonian limits of the
general equations underlying the cluster dynamics and the density and
pressure profiles of the polytropic cluster models.

For a given $\alpha$, the field equations for these general
relativistic cluster models only allow solutions if $\Phi >
\Phi_0(\alpha)$, with $\Phi_0(\alpha)$ an $\alpha$-dependent minimum
value for the potential at the outer boundary. In other words:~for a
given slope of the distribution function, a model cannot be made
arbitrarily compact. The ratio of the minimum outer radius to the
model's Schwarzschild radius is a rising function of $\alpha$,
increasing from $R/R_S \approx 3.6$ for $\alpha=0$ to $R/R_s=\infty$
for $\alpha=3.5$. For less compact models, always two solutions to the
field equations exist:~one with a higher central redshift than the
most compact model and one with a lower central redshift.

The models we constructed, for $\alpha$-values between $0.05$ and
$3.5$, fully cover the first maximum of the fractional binding, where
dynamical instability is expected to set in. This first maximum is
achieved by models which all have a central redshift below $z_c
\approx 0.55$. The most strongly bound model is characterized by
$\alpha=2.75$ and a central redshift $z_c \approx 0.55$. Models with
steeper distribution functions have lower fractional binding energies
than the $\alpha=2.75$ model whereas models with flatter distribution
functions have higher fractional binding energies. In the limit
$\alpha \rightarrow 3.5$, the binding energy and the central redshift
both tend to zero. This indicates that in this limit the distribution
function has become too steep to allow for anything but the Newtonian
solution:~no models with a finite mass exist for $\alpha>3.5$. Hence,
we can conclude that, at least within the context of this family of
models, the Plummer model by necessity is a purely Newtonian
construct.

\section*{Acknowledgements}

The authors wish to thank H. Dejonghe for his insightful suggestions.
This research has been funded by the Interuniversity Attraction Poles
Programme initiated by the Belgian Science Policy Office (IAP P7/08
CHARM).

\label{lastpage}

\end{document}